\documentclass[a4paper,12pt]{article}
\usepackage[T2A]{fontenc}
\usepackage[utf8]{inputenc}
\usepackage[english]{babel}
\usepackage{amsmath}
\usepackage{amsfonts}
\usepackage{amssymb}
\usepackage{wasysym}
\usepackage{amsthm}
\usepackage{tikz}
\usepackage{tikz-cd}
\usetikzlibrary{decorations.markings}
\usetikzlibrary{positioning}
\usetikzlibrary{patterns,patterns.meta}
\usepackage{braids}
\usepackage{array}
\usepackage{ytableau}
\usepackage{longtable}
\usepackage{empheq}
\usepackage{multicol}
\usepackage{mathrsfs}
\usepackage{amssymb}
\usepackage{diagbox}
\usepackage{color}
\usepackage{physics}
\usepackage{ytableau}
\usepackage{cite}
\definecolor{lgreen}{rgb}{0.9,1,0.8}

\usepackage[most]{tcolorbox}

\usepackage{pdflscape}
\usepackage{array, longtable}
\newcolumntype{C}{>{$}c<{$}}

\usepackage{pb-diagram}
\usepackage{setspace}
\usepackage[hidelinks,colorlinks=true,unicode]{hyperref}
\hypersetup{
	linkcolor=magenta,
	citecolor=magenta,
	filecolor=magenta,
	urlcolor=magenta,
}
\usepackage{url}
\usepackage[left=2.3cm,right=2.3cm,top=2.3cm,bottom=2.3cm,bindingoffset=0cm]{geometry}

\def\be{\begin{eqnarray}}
	\def\ee{\end{eqnarray}}

\begin{document}
	\hfill MIPT/TH-16/25
	
	\hfill ITEP/TH-21/25
	
	\hfill IITP/TH-18/25
	
	\vskip 1.5in
	\begin{center}
		
		{\bf\Large Algebra of operators for Q-Schur polynomials}
		
		\vskip 0.2in
		\renewcommand{\thefootnote}{\fnsymbol{footnote}}
		{Nikita Tselousov$^{1,2,3,4}$\footnote[4]{e-mail: tselousov.ns@phystech.edu}}
		\vskip 0.2in
		\renewcommand{\thefootnote}{\roman{footnote}}
		{\small{
				\textit{$^1$MIPT, 141701, Dolgoprudny, Russia}
				\vskip 0 cm
				\textit{$^2$NRC “Kurchatov Institute”, 123182, Moscow, Russia}
				\vskip 0 cm
				\textit{$^3$IITP RAS, 127051, Moscow, Russia}
				\vskip 0 cm
				\textit{$^4$ITEP, Moscow, Russia}
		}}
	\end{center}
	
	\vskip 0.2in
	\baselineskip 16pt
	
	\centerline{ABSTRACT}

	\bigskip
	
	{\footnotesize
		We consider algebras acting on Schur and Q-Schur polynomials, corresponding to Kadomtsev–Petviashvili (KP) and BKP hierarchies. We present them in the spirit of affine Yangians, paying special attention to commutative subalgebras, box additivity property of eigenvalues and single hook expansion of operators.	
	}
	
	\bigskip
	
	\bigskip

	\ytableausetup{boxsize = 0.5em}
	
	\section{Introduction and discussion}
	Integrability is always connected to a hidden symmetry. A large class of these symmetries is described by Yangians \cite{Drinfeld:1985rx, Faddeev:1996iy, Dolan:2004ps, Tsymbaliuk:2014fvq, Okounkov:2015spn, Prochazka:2015deb} and toroidal/DIM algebras \cite{Ding:1996mq, Miki:2007mer, Mironov:2016yue, Awata:2016riz, Feigin2}, that are attracting more and more attention nowadays \cite{Galakhov:2020vyb, Mironov:2019uoy, Mironov:2020pcd, Galakhov:2023mak, Galakhov:2024mbz, Mironov:2024sbc}. Representation theory of these algebras substantially rely on Macdonald theory \cite{Macdonald} and on the notion of crystals \cite{Yamazaki:2010fz, Galakhov:2021xum, NW}, that are generalizations of classical Young diagrams. 
	
	Being a part of the fundamental theory, algebras of hidden symmetries should be defined via few simple initial postulates. The main part of it is, of course, should be integrability itself, that comes down to the commutativity of some operators $\mathcal{O}_a$
	\begin{equation}
		\Big[ \mathcal{O}_a, \mathcal{O}_b \Big] = 0.
	\end{equation}
	The common set of their eigenfunctions distinguishes members of Macdonald family $P_{\lambda}$ among all functions in the proper Hilbert space
	\begin{equation}
		\mathcal{O} \, P_{\Lambda} = \mathcal{E}_{\Lambda} \, P_{\Lambda},
	\end{equation}
	where eigenfunctions $P_{\Lambda}$ and eigenvalues $\mathcal{E}_{\Lambda}$ are enumerated by some crystals $\Lambda$. Integrability implies that the eigenvalues should also be "integrable" in the special sense 
	\begin{equation}
		\mathcal{E}_{\Lambda} = \sum_{\Box \in \Lambda} \Omega_{\Box},
	\end{equation}
	i.e. they should be given by sum over all atoms $\Box$ of the crystal $\Lambda$ of a proper local function $\Omega_{\Box}$. We argue that this postulate indeed leads to meaningful results in two simple cases.

	In this short note we studied the algebra of operators for Schur and Q-Schur polynomials, enumerated by Young-like diagrams. Both Schur $S_{\lambda}$ and Q-Schur $Q_{\lambda}$ polynomials are special limits of Macdonald polynomials\footnote{In our notation Macdonald polynomials satisfy triangular decomposition $M_{\lambda}^{q,t} = m_{\lambda} + \sum_{\mu < \lambda} \, K_{\lambda \mu}^{q,t} \, m_{\mu}$ and Cauchy identity $\sum_{\lambda} \frac{M_{\lambda}^{q,t}(p) \, M_{\lambda}^{q,t}(\bar{p})}{||M_{\lambda}^{q,t}||^2} = \exp \left( \sum_{k = 1} \frac{p_k \, \bar{p}_k}{k} \cdot \frac{t^{2k}-1}{q^{2k}-1}\right)$.}
	\begin{equation}
		S_{\lambda} = \lim_{\substack{q \to 1 , \, t \to 1}} M_{\lambda}^{q,t},
	\end{equation}
	\begin{equation}
		Q_{\lambda} = \lim_{\substack{q \to 0 , \, t \to -\sqrt{-1}}} M_{\lambda}^{q,t}.
	\end{equation}
	In the case of Q-Schur polynomials the set of diagrams enumerating the polynomials is given by strict partitions (SP) -- integer partitions into distinct numbers. The limit $q \to 0, t \to - \sqrt{-1}$ of Macdonald polynomials is well defined even for a non-strict partition, however one should ignore it while considering Q-Schur case.
	
	According to the general method \cite{Galakhov:2025phf} we analyze three types of operators with gradings $1$, $0$ and $-1$. Positive grading operator acts on polynomials by adding a box to the diagrams, negative grading operator removes boxes and zero grading operator acts diagonally. Provided that the eigenvalues are given by the sum of local functions over the diagram -- box additivity property -- the resulting algebra includes obvious commutative family of diagonal operators.
	
	The Schur case corresponds to the algebra $W_{1+\infty}$ \cite{PSR, Awata:1994tf} and KP integrability \cite{Jimbo:1983if}. Schur polynomials arise in Fock representation, which vectors are enumerated by Young diagrams, hence this representation can be realized in the space of polynomials of time variables $p_k$. 
	
	We present the algebra relations in the equivalent form of degenerate affine Yangian $Y(\hat{\mathfrak{gl}}_1)$ with parameters $\epsilon_1 = 1, \epsilon_2 = -1$. The approach to the algebra via Yangian generators $\hat{e}_k, \hat{f}_k, \hat{\psi}_k$ \cite{Drinfeld:1987sy} allows one to analyze different commutative subalgebras that would correspond to quantum integrable systems in a selected realization \cite{Mironov:2020pcd, Mironov:2023zwi}. The first interesting commutative subalgebra is the family of $\hat{\psi}_k$ operators that correspond to Casimir operators \cite{Mironov:2019mah, Mironov:2021taq, Mironov:2021wfh} and features box additivity property of eigenvalues and single hook expansions \cite{Ker1Mironov:2019, Mironov:2019mah}. Another interesting commutative subalgebras include infinite number of operator families parametrized by integer number -- so called integer rays \cite{Mironov:2020pcd}. Commutativity property of integer rays relies only on Serre relations \cite{Mironov:2023wga} and correspond to quantum integrable systems of Calogero type \cite{MMintsystWLZZ}.
	
	We lift selected structures presented in the Schur case to the Q-Schur one, that correspond to BKP integrability \cite{DATE1982343, Alexandrov:2020yxf, Alexandrov:2020nzt, Drachov:2023xyz}. The list of common features of Schur and Q-Schur cases:
	\begin{itemize}
		\item there exist time variables for Young-like diagrams;
		\item the minimal time variable adds and removes one box in the diagrams;
		\item the simplest diagonal operator has finite spin \footnote{By the spin we mean the number of variables and derivatives in the normal ordered form};
		\item all diagonal operators satisfy box additivity property;
		\item there exists recursion relation for eigenvalue functions;
		\item all diagonal operators have restricted expansion in polynomial basis -- like single hook expansion;
		\item there exist quadratic relations of operators with universal coefficients;
		\item there exist Serre-like relations of higher order;
		\item there are integer ray commutative families implied by commutative family of time variables;
		\item the polynomials themselves can be computed via the small set of relations on the Young-like diagrams.
	\end{itemize}
	We observe in Schur and Q-Schur cases that some properties in this list are connected to each other, however in general setup the connection may be violated and some properties may be lost \cite{MThunt, Morozov:2023vra, Wang1, Morozov:2018fga, Qschurs1, Morozov:2020ccp}. Identification of the truly universal properties and connections between them is the task of the future theory that is to be constructed. 
	
	This paper is organized as follows and goes in parallel for Schur and Q-Schur case. In Sections \ref{sec:: Schur basics} and \ref{sec:: Q Schur basics} we discuss basic definitions and formulas. In Sections \ref{sec:: Schur algebra} and \ref{sec::towards the algebra} we discuss the algebras acting on the polynomials. In Sections \ref{sec:: Schur comm integer rays} and \ref{sec:: Q Schur comm rays} we consider integer ray commutative families of operators and provide an algorithm to reconstruct the polynomials from the first commutative families in Sections \ref{sec::Schur from first family} and \ref{sec:: Q Schurs from the first family}. In the last Sections \ref{sec:: Schur Casimirs} and \ref{sec:: Q Schur Casimirs} we develop a general theory of operators with box additivity property and present explicit formulas for all diagonal operators with the help of special recursion relations. 
	
	\section{Schur case}
	\subsection{Basic formulas}
	\label{sec:: Schur basics}
\begin{figure}[h!]
		\begin{equation*}
			\begin{tikzpicture}[scale=0.35]
				\fill[pattern=north east lines] (0,0) rectangle (16,-0.5);
				\fill[pattern=north east lines] (-0.5,-0.5) rectangle (0, 13);
				\draw[line width=0.5pt](0,0)-- (16,0); 
				\draw[line width=0.5pt](0,0)-- (0,13);
				\foreach \i/\j in {0/0, 0/1, 0/2, 0/3, 0/4, 0/5, 0/6, 0/7, 0/8, 0/9, 0/10, 1/0, 1/1, 1/2, 1/3, 1/4, 1/5, 1/6, 1/7, 1/8, 1/9, 2/0, 2/1, 2/2, 2/3, 2/4, 2/5, 2/6, 2/7, 3/0, 3/1, 3/2, 3/3, 3/4, 3/5, 4/0, 4/1, 4/2, 4/3, 4/4, 4/5, 5/0, 5/1, 5/2, 5/3, 5/4, 6/0, 6/1, 6/2, 6/3, 7/0, 7/1, 7/2, 8/0, 8/1, 8/2, 9/0, 9/1, 10/0, 11/0, 12/0}
				{
					\draw[thick] (\i,\j) -- (\i+1,\j) -- (\i+1,\j+1) -- (\i,\j+1) -- (\i,\j);
				}
				\filldraw [gray] 
				(0.5,11.5) circle (4pt)
				(1.5,10.5) circle (4pt)
				(2.5,8.5) circle (4pt)
				(3.5,6.5) circle (4pt)
				(5.5,5.5) circle (4pt)
				(6.5,4.5) circle (4pt)
				(7.5,3.5) circle (4pt)
				(9.5,2.5) circle (4pt)
				(10.5,1.5) circle (4pt)
				(13.5,0.5) circle (4pt);
				\draw 
				(0.5,10.5) node {$\times$}
				(1.5,9.5) node {$\times$}
				(2.5,7.5) node {$\times$}
				(4.5,5.5) node {$\times$}
				(5.5,4.5) node {$\times$}
				(6.5,3.5) node {$\times$}
				(8.5,2.5) node {$\times$}
				(9.5,1.5) node {$\times$}
				(12.5,0.5) node {$\times$};
			\end{tikzpicture}
		\end{equation*}
		\caption{Example of Young diagram $\lambda = [13,10,9,7,6,5,3,3,2,2,1]$. In our notation $\lambda_i$ is the length of $i$-th row counting from the bottom. Young diagram $\lambda$ can be considered as a way of tight packing of $|\lambda| = \sum_{i} \lambda_i$ identical boxes in the corner. Gray dots correspond to the set of possible positions for the new boxes -- i.e. $\text{Add}(\lambda)$. By the sign "$\times$" we mark boxes that can be removed from the diagram $\lambda$, i.e. the set $\text{Rem}(\lambda)$. }
		\label{fig::YD example}
	\end{figure}
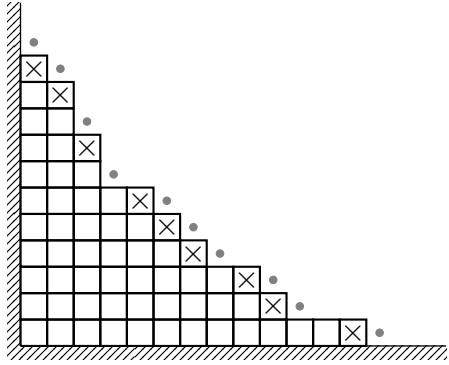

	Schur polynomials $S_{\lambda}(x)$  form a distinguished basis in the space of symmetric polynomials of $x_i$ variables, where $i = 1, \ldots, N$. These polynomials are enumerated by integer partitions $\lambda = [\lambda_1, \lambda_2, \ldots, \lambda_{l(\lambda)}]$, where $\lambda_1 \geqslant \lambda_2 \geqslant  \ldots \geqslant \lambda_{l(\lambda)}$ and $\lambda_i \in \mathbb{Z}_{+}$. The number of partitions of a given size can be seen from the following well-known generating function
	\begin{equation}
		\prod_{k=1} \frac{1}{1-x^{k}} = 1+x+2 x^2+3 x^3+5 x^4+7 x^5+11 x^6+15 x^7+22 x^8 + \ldots \, .
	\end{equation}
	Integer partitions can be represented as Young diagrams (see Fig. \ref{fig::YD example}).
	Schur polynomials are characters of $GL(N)$ groups and therefore can be computed via Weyl determinant formula
	\begin{equation}
		S_{\lambda}(x) = \frac{\det \left( x^{\lambda_j + N - j}_i \right) }{ \det \left( x^{N - j}_i \right) }.
	\end{equation}
	In our presentation we use Schur polynomials in terms of time variables $p_k$, where $k = 1, 2, 3, \ldots$ and the change of variables is given by the Miwa transformation
	\begin{equation}
		p_k = \sum_{i = 1}^{N} (x_i)^k .
	\end{equation}
	Schur polynomials $S_{\lambda}(p)$ are homogeneous polynomials provided that the grading of variable $p_k$ is $k$. Explicit examples of Schur polynomials for small levels are presented in Sec.\ref{sec:: Q Schurs from the first family}.

	\subsection{$W_{1+\infty}$ algebra as degenerate  affine Yangian $\mathfrak{gl}_1$}
	\label{sec:: Schur algebra}
	For applications to the representation theory of infinite-dimentional algebras the following properties of Schur polynomials are needed. The first is famous Pieri rule \cite{Macdonald}
	\begin{equation}
		\label{Pieri add Schur}
		p_1 \cdot S_{\lambda} = \sum_{\Box \in \text{Add}(\lambda)} S_{\lambda + \Box}.
	\end{equation}
	We denote as $\text{Add}(\lambda)$ the set of possible positions of the box $\Box$, such that the resulting diagram $\lambda + \Box$ is a proper Young diagram. Dual Pieri rule has similar form
	\begin{equation}
		\label{Pieri rem Schur}
		\frac{\partial}{\partial p_1} S_{\lambda} = \sum_{\Box \in \text{Rem}(\lambda)} S_{\lambda - \Box},
	\end{equation}
	where the definition of the set $\text{Rem}(\lambda)$ is clear from the Fig.\ref{fig::YD example}. Another important property of Schur polynomials is that they are eigenfunctions of simple cut-and-join operator $\hat{W}$ \cite{MMN} 
	\begin{equation}
		\hat{W} = \frac{1}{2} \sum_{a, b = 1}^{\infty}  (a+b) \, p_a p_b \frac{\partial }{\partial p_{a+b}} + a b \, p_{a+b} \frac{\partial}{\partial p_a} \frac{\partial}{\partial p_b},
	\end{equation}
	\begin{equation}
		\hat{W} S_{\lambda} = \kappa_{\lambda} \, S_{\lambda} \, .
	\end{equation}
	The main feature of the above operator is that the eigenvalues $\kappa_{\lambda}$ are given by the sum of local quantities over all boxes in the diagram $\lambda$
	\begin{equation}
		\kappa_{\lambda} = \sum_{\Box \in \lambda} j_{\Box} - i_{\Box}.
	\end{equation}
	In our notation box $\Box$ has horizontal and vertical coordinates $j_{\Box}$ and $i_{\Box}$ respectively. We call this property of eigenvalues \textit{box additivity}. Using three operators:
	\begin{itemize}
		\item $p_1$ -- adds box to the Young diagram,
		\item $\frac{\partial}{\partial p_1}$ -- removes box from the Young diagram,
		\item $\hat{W}$ -- diagonal operator with box additivity property,
	\end{itemize}
	one can construct an infinite family of  commutative operators, that act diagonally on Schur polynomials. For this purpose we introduce auxiliary operators $\hat{e}_k, \hat{f}_k$ that are defined by the recursive relations
	\begin{equation}
		\hat{e}_k = \Big[ \hat{W}, \hat{e}_{k-1} \Big] \hspace{20mm} \hat{e}_0 = p_1 , 
	\end{equation}
	\begin{equation}
		\hat{f}_k = -\Big[ \hat{W}, \hat{f}_{k-1} \Big] \hspace{20mm} \hat{f}_0 =  -\frac{\partial}{\partial p_1}.
	\end{equation}
	These operators act on Schur polynomials by adding and removing boxes but with local coefficients
	\begin{equation}
		\hat{e}_k \, S_{\lambda} = \sum_{\Box \in \text{Add}(\lambda)} (j_{\Box} - i_{\Box})^k \cdot S_{\lambda + \Box},
		\label{higher e action}
	\end{equation}
	\begin{equation}
		\hat{f}_k \, S_{\lambda} = (-1) \sum_{\Box \in \text{Rem}(\lambda)} (j_{\Box} - i_{\Box})^k \cdot S_{\lambda - \Box}.
	\end{equation}
	Commutators of $\hat{e}_k, \hat{f}_k$ operators
	\begin{equation}
		\hat{\psi}_{k+l} = \Big[ \hat{e}_{k}, \hat{f}_{k} \Big] 
	\end{equation}
	form  commutative family
	\begin{equation}
		\Big[ \hat{\psi}_a, \hat{\psi}_b \Big] = 0.
	\end{equation}
	Commutativity of $\hat{\psi}_a$ operators is a direct consequence of box additivity property of $\hat{W}$ and of the fact, that initial operator $\Big[ \hat{e}_0, \hat{f}_0 \Big] = -\Big[ p_1, \frac{\partial}{\partial p_1} \Big] = 1$ is diagonal on Schur polynomials \cite{Galakhov:2025phf}. The initial operator $\hat{W}$ is included in the  commutative family 
	\begin{equation}
		\hat{\psi}_3 = 6 \hat{W},
	\end{equation}
	therefore Schur polynomials form the set of common eigenfunctions of operators $\hat{\psi}_a$.
	
	It is not obvious in our presentation, however the set of operators $\hat{e}_a, \hat{f}_a, \hat{\psi}_a$, $a = 0, 1, 2, \ldots$, generates infinite-dimensional Lie algebra $W_{1+\infty}$  in Fock representation. We provide relations of $W_{1+\infty}$ algebra in the equivalent form of degenerate affine Yangian $Y(\hat{\mathfrak{gl}}_1)$ \cite{Prochazka:2015deb, Tsymbaliuk:2014fvq}, where parameters are set $\epsilon_1 = 1, \epsilon_2 = -1$ (i.e. $\sigma = -1, \sigma_3 = 0$)
	\begin{align}
		\begin{aligned}
			\Big[ \hat{\psi}_n, \hat{\psi}_m \Big] &= 0, \\
			\Big[ \hat{e}_n, \hat{f}_m \Big]  &= \hat{\psi}_{n+m},
		\end{aligned}
	\end{align}
	\begin{align}
		\begin{aligned}
			\Big[\hat{\psi}_0, \hat{e}_n \Big] &= 0 &\hspace{15mm} \Big[\hat{\psi}_1, \hat{e}_n \Big] &= 0 &\hspace{15mm} \Big[\hat{\psi}_2, \hat{e}_n \Big] &= 2 \hat{e}_n\\
			\Big[\hat{\psi}_0, \hat{f}_n \Big] &= 0 &\hspace{15mm} \Big[\hat{\psi}_1, \hat{f}_n \Big] &= 0 &\hspace{15mm} \Big[\hat{\psi}_2, \hat{f}_n \Big] &= -2 \hat{f}_n\\
		\end{aligned}
	\end{align}
	\begin{align}
		\begin{aligned}
			\Big[ \hat{e}_{n+3}, \hat{e}_{m} \Big] - 3 \Big[ \hat{e}_{n+2}, \hat{e}_{m+1} \Big] + 3 \Big[ \hat{e}_{n+1}, \hat{e}_{m+2} \Big] - \Big[ \hat{e}_{n}, \hat{e}_{m+3} \Big] -  \Big[ \hat{e}_{n+1}, \hat{e}_{m} \Big] + \Big[ \hat{e}_{n}, \hat{e}_{m+1} \Big] = 0 \\
			\Big[ \hat{f}_{n+3}, \hat{f}_{m} \Big] - 3 \Big[ \hat{f}_{n+2}, \hat{f}_{m+1} \Big] + 3 \Big[ \hat{f}_{n+1}, \hat{f}_{m+2} \Big] - \Big[ \hat{f}_{n}, \hat{f}_{m+3} \Big] -  \Big[ \hat{f}_{n+1}, \hat{f}_{m} \Big] + \Big[ \hat{f}_{n}, \hat{f}_{m+1} \Big] = 0 
		\end{aligned}
	\\
		\begin{aligned}
			\label{Schur relations psi e} 
			\Big[ \hat{\psi}_{n+3}, \hat{e}_{m} \Big] - 3 \Big[ \hat{\psi}_{n+2}, \hat{e}_{m+1} \Big] + 3 \Big[ \hat{\psi}_{n+1}, \hat{e}_{m+2} \Big] - \Big[ \hat{\psi}_{n}, \hat{e}_{m+3} \Big] -  \Big[ \hat{\psi}_{n+1}, \hat{e}_{m} \Big] + \Big[ \hat{\psi}_{n}, \hat{e}_{m+1} \Big] = 0 \\
			\Big[ \hat{\psi}_{n+3}, \hat{f}_{m} \Big] - 3 \Big[ \hat{\psi}_{n+2}, \hat{f}_{m+1} \Big] + 3 \Big[ \hat{\psi}_{n+1}, \hat{f}_{m+2} \Big] - \Big[ \hat{\psi}_{n}, \hat{f}_{m+3} \Big] -  \Big[ \hat{\psi}_{n+1}, \hat{f}_{m} \Big] + \Big[ \hat{\psi}_{n}, \hat{f}_{m+1} \Big] = 0 
		\end{aligned}
	\end{align}
	\begin{align}
		\begin{aligned}
			\label{serre rel}
			\text{Sym}_{i,j,k} \Big[\hat{e}_i, \Big[ \hat{e}_j, \hat{e}_{k+1} \Big] \Big] =0 \\
			\text{Sym}_{i,j,k} \Big[\hat{f}_i, \Big[ \hat{f}_j, \hat{f}_{k+1} \Big] \Big] =0 
		\end{aligned}
	\end{align}

	\subsection{Commutative integer rays of operators}
	\label{sec:: Schur comm integer rays}
	Algebra $W_{1+\infty}$ includes infinite families of  commutative operators enumerated by integer numbers \cite{Mironov:2020pcd}. Commutativity property of these families follows from cubic Serre relations \eqref{serre rel} and persists in any representation \cite{Mironov:2023wga}, while we consider only Fock representation. We discuss  commutative families in subalgebra generated by $\hat{e}_k$ operators, while the other families corresponding to $\hat{f}_k$ operators are constructed in the similar way. The first family $\hat{H}_a^{(0)}$ is defined by the following formulas
	\begin{equation}
		\hat{H}_a^{(0)} = \frac{1}{a} \Big[ \hat{e}_1, \hat{H}_{a-1}^{(0)} \Big], \hspace{20mm} \hat{H}_0^{(0)} = \hat{e}_0.
	\end{equation}
	In Fock representation these operators correspond to time variables
	\begin{equation}
		\hat{H}_a^{(0)} = p_{a+1}.
	\end{equation}
	Higher families $\hat{H}_a^{(M)}$ are defined in the following way
	\begin{equation}
		\hat{H}_a^{(M)} = \frac{1}{a} \Big[ \hat{e}_{M+1}, \hat{H}_{a-1}^{(M)} \Big], \hspace{20mm} \hat{H}_0^{(M)} = \hat{e}_M.
	\end{equation}
	\begin{equation}
		\Big[ \hat{H}_a^{(M)}, \hat{H}_b^{(M)} \Big] = 0
	\end{equation}
	Operators $\hat{H}_a^{(1)}$ of the family $M=1$ play a central role in the theory of  WLZZ models (negative branch) \cite{Wang:2022fxr, MMMPWZ, MMMPZ}. These  commutative families are connected to Hamiltonians of Calogero type integrable systems \cite{Mironov:2023zwi}.
	\subsection{Schur polynomials from the first commutative family}
	\label{sec::Schur from first family}
	Schur polynomials $S_{\lambda}$ can be encoded in an elegant way by the following relations:
	\begin{itemize}
		\item Operators $\hat{e}_0, \hat{e}_1$ add boxes \eqref{higher e action}
		\begin{equation}
			\hat{e}_0 \, S_{\lambda} = \sum_{\Box \in \text{Add}(\lambda)} S_{\lambda + \Box},
		\end{equation}
		\begin{equation}
			\hat{e}_1 \, S_{\lambda} = \sum_{\Box \in \text{Add}(\lambda)} (j_{\Box}-i_{\Box}) \cdot S_{\lambda + \Box}.
		\end{equation}
		\item Commutative operators $\hat{H}_{a}^{(0)}$ correspond to time variables
		\begin{equation}
			p_{a+1} = \hat{H}_{a}^{(0)} = \frac{1}{a!} \underbrace{\Big[ \hat{e}_1 ,\Big[ \hat{e}_1, \ldots, \Big[ \hat{e}_1}_{a}, \hat{e}_0 \Big] \ldots \Big] \Big]  
		\end{equation}
	\end{itemize}
	It is not needed to know explicit from of operator $\hat{e}_1$ in terms of time variables, however one can extract Schur polynomials from the above data. In the beginning we postulate $S_{\varnothing} = 1$, then higher polynomials can be computed. We show explicit examples from small levels. 
	\begin{equation}
		p_1 = \hat{e}_0, \hspace{10mm} p_2 = \Big[ \hat{e}_1, \hat{e}_0 \Big], \hspace{10mm} p_3 = \frac{1}{2} \Big[\hat{e}_1, \Big[ \hat{e}_1, \hat{e}_0 \Big] \Big]
	\end{equation}

	2 level:
	
	\begin{align}
		\begin{aligned}
			p_1^2 \cdot 1 &= \hat{e}_0 \hat{e}_0 \, S_{\varnothing} = S_{[2]} + S_{[1,1]} \\
			p_2 \cdot 1 &= \Big[ \hat{e}_1, \hat{e}_0 \Big] \, S_{\varnothing} = S_{[2]} - S_{[1,1]}
		\end{aligned}
	\end{align}
	Solving this simple linear system we derive
	\begin{equation}
		S_{[2]} = \frac{p_1^2+p_2}{2}, \hspace{10mm} S_{[1,1]} = \frac{p_1^2-p_2}{2}.
	\end{equation}

	3 level:
	
	\begin{align}
		\begin{aligned}
			p_1^3 \cdot 1 &= \hat{e}_0 \hat{e}_0 \hat{e}_0 \, S_{\varnothing} = S_{[3]} + 2 S_{[2,1]} + S_{[1,1,1]} \\
			p_2 p_1 \cdot 1 &= \hat{e}_0 \Big[ \hat{e}_1, \hat{e}_0 \Big] \, S_{\varnothing} = S_{[3]} - S_{[1,1,1]} \\
			p_3 \cdot 1 &= \frac{1}{2} \Big[ \hat{e}_1, \Big[ \hat{e}_1, \hat{e}_0 \Big] \Big] \, S_{\varnothing} = S_{[3]} -S_{[2,1]} + S_{[1,1,1]}
		\end{aligned}
	\end{align}
	From simple linear system
	\begin{equation}
		S_{[3]} = \frac{1}{6} \left(p_1^3+3 p_1 p_2+2 p_3\right), \hspace{10mm} S_{[2,1]} = \frac{1}{3} \left(p_1^3-p_3\right), \hspace{10mm} S_{[1,1,1]} = \frac{1}{6} \left(p_1^3-3 p_1 p_2+2 p_3\right).
	\end{equation}
	This approach to Schur polynomials does not refer to determinant formulas, however it uses the geometry of Young diagrams and function $j_{\Box} - i_{\Box}$, that makes it applicable to other cases if one changes the geometry and the function. 
	\subsection{Explicit form of operators $\hat{\psi}_a$}
	\label{sec:: Schur Casimirs}
	For $W_{1 + \infty}$ in Fock representation \textit{all} operators \footnote{Except cases $a = 0,1$, where $\hat{\psi}_0 = 1, \hat{\psi}_1 = 0$.} $\hat{\psi}_a$ have box additivity property (this property is violated after $\beta$-deformation \cite{Prochazka:2015deb}). Therefore we develop a general theory of operators with box additivity property for Schur polynomials. Consider the following diagonal operator
	\begin{equation}
		\hat{\psi}_{\omega} \, S_{\lambda} = \left( \sum_{\Box \in \lambda} \omega(j_{\Box} - i_{\Box}) \right) S_{\lambda},
	\end{equation}
	where $\omega(x)$ is an arbitrary function that we treat as a parameter of the above operator. This operator has simple form in terms of Schur polynomials and corresponding dual operators. In other words, we represent the operator in the following basis with some coefficients $A_{\mu, \nu}$
	\begin{equation}
		\hat{\psi}_{\omega} = \sum_{\mu, \nu} A_{\mu, \nu} \, S_{\mu} \hat{S}_{\nu^{T}},
	\end{equation}
	where the sum runs over all Young diagrams $\mu,\nu$ of equal size $|\mu| = |\nu|$. $\nu^T$ means transposed Young diagram and dual operators $\hat{S}_{\lambda}$ are defined in the following way
	\begin{equation}
		\hat{S}_{\lambda} := S_{\lambda} \left( p_k \to k \frac{\partial}{\partial p_k} \right).
	\end{equation}
	Operators with box additivity property have additional \textit{special property} -- their expansion involve only \textit{single hook} Young diagrams, that we denote for simplicity
	\begin{equation}
		(k|n) := [n - k + 1, \underbrace{1, \ldots, 1}_{k-1}] = [n - k + 1, 1^{k-1}].
	\end{equation}
	We leave the study of origins of this mysterious single hook constraints for future study.
	Here $n$ is the size of the diagram and $k$ is the number of rows. Then the resulting formula for the operator reads with arbitrary function $\omega(x)$
	\begin{tcolorbox}
	\begin{equation}
		\hat{\psi}_{\omega} = \sum_{n = 1}^{\infty} \sum_{i,j = 1}^{n} (-1)^{n + 1+ i + j} \cdot \omega(j - i) \cdot S_{(i|n)} \hat{S}_{(j|n)^{T}}.
	\end{equation}
	\end{tcolorbox}
	Then the description of operators $\hat{\psi}_a$ comes down to the description of functions $\omega_a(x)$
	\begin{equation}
		\hat{\psi}_a \, S_{\lambda} = \left( \sum_{\Box \in \lambda} \omega_{a}(j_{\Box} - i_{\Box})\right) S_{\lambda}.
	\end{equation}
	From several explicit examples
	\begin{align}
		\begin{aligned}
			\omega_2(x) &= 2,\\
			\omega_3(x) &= 6x,\\
			\omega_4(x) &= 12x^2+2,\\
			\omega_5(x) &= 20 x^3+10 x,\\
			\omega_6(x) &= 30 x^4+30 x^2+2,\\
			\omega_7(x) &= 42 x^5+70 x^3+14 x,\\
			\omega_8(x) &= 56 x^6+140 x^4+56 x^2+2,\\
			\ldots \\
		\end{aligned}
	\end{align}
	one can deduce the general formula
	\begin{equation}
		\label{Schur omega}
		\boxed{\boxed{
		\omega_a(x) = (x+1)^a + (x-1)^a -2 x^a.
	}}
	\end{equation}
	Explicit formula for $\hat{\psi}_a$ generators for $a = 0,1,2,\ldots$ in Fock representation then follows
	\begin{tcolorbox}
		\begin{equation}
			\hat{\psi}_{a} = \delta_{a,0} + \sum_{n = 1}^{\infty} \sum_{i,j = 1}^{n} (-1)^{n + 1+ i + j} \cdot \Big[ (j-i+1)^a + (j-i-1)^a -2 (j-i)^a \Big] \cdot S_{(i|n)} \hat{S}_{(j|n)^{T}}.
		\end{equation}
	\end{tcolorbox}
	We verify our formula \eqref{Schur omega} with the help of commutational relations of the $W_{1+\infty}$. We use a general fact about an operator with box additivity property
	\begin{equation}
		\Big[ \hat{\psi}_a, \hat{e}_k \Big] S_{\lambda} = \sum_{\Box \in \text{Add}(\lambda)} \omega_a (j_{\Box}-i_{\Box}) \cdot (j_{\Box}-i_{\Box})^k \, S_{\lambda + \Box}.
	\end{equation}
	Then relations \eqref{Schur relations psi e} impose the following constraint for $\omega_a(x)$
	\begin{equation}
		\omega_{a+3}(x) - 3 x \cdot \omega_{a+2}(x) + \left( 3 x^2 - 1 \right) \cdot \omega_{a+1}(x)  +(x - x^3) \cdot \omega_{a}(x)  = 0,
	\end{equation}
	that is indeed satisfied by \eqref{Schur omega}. This recursive relation is a key to understand the functions $\omega_a(x)$. It is linear and has degree 3 therefore it has three solutions
	\begin{equation}
		\omega_a^{(1)}(x) = (x+1)^a, \hspace{20mm} \omega_a^{(1)}(x) = (x-1)^a, \hspace{20mm} \omega_a^{(1)}(x) = (x)^a,
	\end{equation}
	due to the characteristic polynomial
	\begin{equation}
		t^{a+3} - 3 x \cdot t^{a+2} + \left( 3 x^2 - 1 \right) \cdot t^{a+1}  +(x - x^3) \cdot t^{a} = t^a (t-1-x ) (t-x) (t-x+1).
	\end{equation}
	These three solutions are also follow from structure function of the affine Yangian $\mathfrak{gl}_1$ \cite{Prochazka:2015deb} 
	\begin{equation}
		\varphi(x) = \frac{(x+\epsilon_1)(x+\epsilon_2)(x-\epsilon_1-\epsilon_2)}{(x-\epsilon_1)(x-\epsilon_2)(x+\epsilon_1+\epsilon_2)},
	\end{equation}
	for special value of parameters $\epsilon_1 = 1, \epsilon_2 = -1$.

	\section{Q-Schur case}
	\subsection{Basic formulas}
	\label{sec:: Q Schur basics}
\begin{figure}[h!]
		\begin{equation*}
			\begin{tikzpicture}[scale=0.35]
				\foreach \i/\j in {0/0, 1/0, 2/0, 3/0, 4/0, 5/0, 6/0, 7/0, 8/0, 9/0, 10/0, 11/0, 12/0, 13/0, 14/0, 1/1, 2/1, 3/1, 4/1, 5/1, 6/1, 7/1, 8/1, 9/1, 10/1, 11/1, 12/1, 2/2, 3/2, 4/2, 5/2, 6/2, 7/2, 8/2, 9/2, 10/2, 11/2, 3/3, 4/3, 5/3, 6/3, 7/3, 8/3, 9/3, 4/4, 5/4, 6/4, 7/4, 5/5, 6/5, 7/5, 6/6, 7/6}
				{
					\draw[thick] (\i,\j) -- (\i+1,\j) -- (\i+1,\j+1) -- (\i,\j+1) -- (\i,\j);
				}
				\fill[pattern=north east lines] (0,0) rectangle (18,-0.5);
				\fill[pattern=north east lines] (-0.5,-0.5) rectangle (0, 1.0);
				\fill[pattern=north east lines] (0.5, 1.0) rectangle (1.0, 2.0);
				\fill[pattern=north east lines] (1.5, 2.0) rectangle (2.0, 3.0);
				\fill[pattern=north east lines] (2.5, 3.0) rectangle (3.0, 4.0);
				\fill[pattern=north east lines] (3.5, 4.0) rectangle (4.0, 5.0);
				\fill[pattern=north east lines] (4.5, 5.0) rectangle (5.0, 6.0);
				\fill[pattern=north east lines] (5.5, 6.0) rectangle (6.0, 7.0);
				\fill[pattern=north east lines] (6.5, 7.0) rectangle (7.0, 8.0);
				\fill[pattern=north east lines] (7.5, 8.0) rectangle (8.0, 9.0);
				\fill[pattern=north east lines] (8.5, 9.0) rectangle (9.0, 10.0);
				\fill[pattern=north east lines] (9.5, 10.0) rectangle (10.0, 11.0);
				\fill[pattern=north east lines] (-0.5, 1.0) rectangle (0.5, 1.5);
				\fill[pattern=north east lines] (0.5, 2.0) rectangle (1.5, 2.5);
				\fill[pattern=north east lines] (1.5, 3.0) rectangle (2.5, 3.5);
				\fill[pattern=north east lines] (2.5, 4.0) rectangle (3.5, 4.5);
				\fill[pattern=north east lines] (3.5, 5.0) rectangle (4.5, 5.5);
				\fill[pattern=north east lines] (4.5, 6.0) rectangle (5.5, 6.5);
				\fill[pattern=north east lines] (5.5, 7.0) rectangle (6.5, 7.5);
				\fill[pattern=north east lines] (6.5, 8.0) rectangle (7.5, 8.5);
				\fill[pattern=north east lines] (7.5, 9.0) rectangle (8.5, 9.5);
				\fill[pattern=north east lines] (8.5, 10.0) rectangle (9.5, 10.5);
				\draw[line width=0.5pt](0,0)-- (18,0); 
				\draw[line width=0.5pt](0,0)-- (0,1);
				\draw[line width=0.5pt](1,1)-- (1,2);
				\draw[line width=0.5pt](2,2)-- (2,3);
				\draw[line width=0.5pt](3,3)-- (3,4);
				\draw[line width=0.5pt](4,4)-- (4,5);
				\draw[line width=0.5pt](5,5)-- (5,6);
				\draw[line width=0.5pt](6,6)-- (6,7);
				\draw[line width=0.5pt](7,7)-- (7,8);
				\draw[line width=0.5pt](8,8)-- (8,9);
				\draw[line width=0.5pt](9,9)-- (9,10);
				\draw[line width=0.5pt](10,10)-- (10,11);
				\draw[line width=0.5pt](0,1)-- (1,1);
				\draw[line width=0.5pt](1,2)-- (2,2);
				\draw[line width=0.5pt](2,3)-- (3,3);
				\draw[line width=0.5pt](3,4)-- (4,4);
				\draw[line width=0.5pt](4,5)-- (5,5);
				\draw[line width=0.5pt](5,6)-- (6,6);
				\draw[line width=0.5pt](6,7)-- (7,7);
				\draw[line width=0.5pt](7,8)-- (8,8);
				\draw[line width=0.5pt](8,9)-- (9,9);
				\draw[line width=0.5pt](9,10)-- (10,10);
				\filldraw [gray] 
				(7.5,7.5) circle (4pt)
				(8.5,4.5) circle (4pt)
				(10.5,3.5) circle (4pt)
				(12.5,2.5) circle (4pt)
				(13.5,1.5) circle (4pt)
				(15.5,0.5) circle (4pt);
				\draw 
				(7.5,6.5) node {$\times$}
				(9.5,3.5) node {$\times$}
				(11.5,2.5) node {$\times$}
				(12.5,1.5) node {$\times$}
				(14.5,0.5) node {$\times$};
			\end{tikzpicture}
		\end{equation*}
		\caption{Example of ladder Young diagram $\lambda = [15,12,10,7,4,3,2]$. In our notation $\lambda_i$ is the length of $i$-th row counting from the bottom. Ladder Young diagram $\lambda$ can be considered as a way of tight packing of $|\lambda| = \sum_{i} \lambda_i$ identical boxes \textit{under an infinite ladder}. Gray dots correspond to the set of possible positions for the new boxes, i.e. $\text{Add}(\lambda)$. By the sign "$\times$" we mark boxes that can be removed from the diagram $\lambda$, i.e. the set $\text{Rem}(\lambda)$. }
		\label{fig::SP example}
	\end{figure}
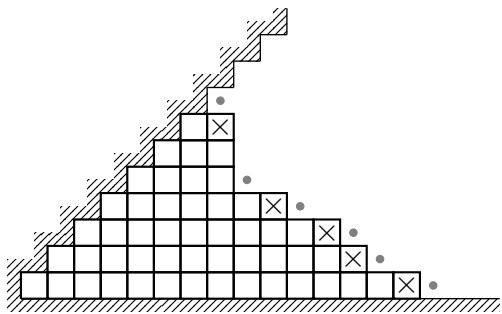

	Q-Schur polynomials $Q_{\lambda}$ are enumerated by strict partitions $\lambda = [\lambda_1, \lambda_2, \ldots, \lambda_{l(\lambda)}]$, where $\lambda_1 > \lambda_2 > \ldots > \lambda_{l(\lambda)}$ and $\lambda_i \in \mathbb{Z}_{+}$. The number of strict partitions of a given size can be seen from the following generating function
	\begin{equation}
		\prod_{k=1} \frac{1}{1-x^{2k-1}} = 1+x+x^2+2 x^3+2 x^4+3 x^5+4 x^6+5 x^7+6 x^8 + \ldots.
	\end{equation}
	We represent strict partitions as a ladder Young diagrams \cite{Azheev:2025wti} (see Fig. \ref{fig::SP example}). 
	One can compute polynomials $Q_{\lambda}$ via the following two step procedure. On the first step an auxiliary polynomials $P_{a,b}$ are computed from the generating function
	\begin{equation}
		\left( \exp\Big\{ 2 \cdot \sum_{k=0}^{\infty} \tau_{k} (z_1^{2k+1}+ z_2^{2k+1})\Big\} - 1\right)  \frac{z_1-z_2}{z_1+z_2} = \sum_{a,b}^{\infty} z_1^a z_2^b \, P_{a,b} \,.
	\end{equation}
	On the second step Q-Schur polynomial for strict partition $\lambda$ \footnote{If number of rows $l(\lambda)$ is odd one should add a zero entry and consider partition $[\lambda_1, \lambda_2, \ldots, \lambda_{l(\lambda)}, 0]$.} is computed as follows
	\begin{equation}
		Q_{\lambda} = \frac{1}{2^{l(\lambda)}} \, \sqrt{\det P_{\lambda_{i}, \lambda_{j}}}.
	\end{equation}
	The resulting polynomials $Q_{\lambda}$ are graded polynomials in variables $\tau_k$ (where $k = 0, 1,2,3, \ldots$) with respect to gradings $[\tau_k] = 2k+1$. Variables $\tau_k$ are relatives of $p_k$ time variables that, for example, is clear from relation on single row polynomials
	\begin{equation}
		\frac{1}{2^{n-1}} \, Q_{[n]} \left( \tau_k \to \frac{4^k \, p_{2k+1}}{2k+1} \right) = S_{[n]} \left( p_{2k} \to 0\right) \,. 
	\end{equation}
	Examples of Q-Schur polynomials for small levels are presented in Sec.\ref{sec:: Q Schurs from the first family}.	
	
	\subsection{Towards the algebra}
	\label{sec::towards the algebra}
	To construct an algebra for Q-Schur polynomials we proceed in a similar way to usual Schur polynomials. We consider operators that add and remove boxes in ladder Young diagrams: multiplication by variable $\tau_0$ gives the simplest Pieri rule
	\begin{equation}
		\tau_0 \cdot Q_{\lambda} = \sum_{\Box \in \text{Add}(\lambda)} Q_{\lambda + \Box} \,, 
	\end{equation}
	while the corresponding derivative $\frac{\partial}{\partial \tau_0}$ gives the dual Pieri rule
	\begin{equation}
		\frac{\partial}{\partial \tau_0} \, Q_{\lambda} = \sum_{\Box \in \text{Rem}(\lambda)} (2 - \delta_{i_{\Box}, j_{\Box}}) \cdot Q_{\lambda - \Box} \,.
	\end{equation}
	Note that in contrast to formulas \eqref{Pieri add Schur}, \eqref{Pieri rem Schur} the coefficient in the r.h.s. of the above formula depends on the position of the removed box. Here $i$ and $j$ are vertical and horizontal coordinates respectively. The main part of the construction of the algebra is the following operator with box additivity property \cite{MMN, Alexandrov:2020yxf, Mironov:2021taq, Azheev:2025wti}
	\begin{align}
		\label{Ham for Q}
		\begin{aligned}
			\hat{U}=\sum_{a,b,c=0}^{\infty}(2(a+b+c)+3)\tau_{a+b+c+1}\frac{\partial^3}{\partial\tau_a\partial\tau_b\partial\tau_c}+ \sum_{a+b=c+d}3(2a+1)(2b+1)\tau_a\tau_b\frac{\partial^2}{\partial\tau_c\partial\tau_d} \\
			\sum_{a,b,c=0}^{\infty}4(2a+1)(2b+1)(2c+1)\tau_a\tau_b\tau_c\frac{\partial}{\partial\tau_{a+b+c+1}}+ \sum_{a=0}^{\infty}(2 a+1) \left(2 a^2+2 a+1\right)\tau_a\frac{\partial}{\partial\tau_a} \, ,
		\end{aligned}
	\end{align}
	that acts diagonally on Q-Schur polynomials
	\begin{equation}
		\hat{U} \, Q_{\lambda} = \Big( \sum_{\Box \in \lambda} (j_{\Box} - i_{\Box}+1)^3 - (j_{\Box} - i_{\Box})^3 \Big) \, Q_{\lambda} \, .
	\end{equation}
	For simplicity we denote $\gamma_{\Box} = (j_{\Box} - i_{\Box}+1)^3 - (j_{\Box} - i_{\Box})^3 = 3(j_{\Box} - i_{\Box})(j_{\Box} - i_{\Box} + 1) + 1$. 
	The algebra that acts on Q-Schur polynomials is generated by the generators $\hat{E}_k, \hat{F}_k, \hat{\Psi}_k$, where $k = 0, 1, 2, \ldots$. We define these generators by the following formulas
	\begin{equation}
		\hat{E}_k = \Big[ \hat{U}, \hat{E}_{k-1} \Big], \hspace{20mm} \hat{E}_0 = \tau_0,
	\end{equation}
	\begin{equation}
		\hat{F}_k = -\Big[ \hat{U}, \hat{F}_{k-1} \Big], \hspace{20mm} \hat{F}_0 = - \frac{\partial}{\partial \tau_0}.
	\end{equation}
	Higher operators $\hat{E}_k, \hat{F}_k$ add and remove boxes
	\begin{equation}
		\hat{E}_k \, Q_{\lambda} = \sum_{\Box \in \text{Add}(\lambda)} (\gamma_{\Box})^k \, Q_{\lambda + \Box} \,, 
	\end{equation}
	\begin{equation}
		\hat{F}_k \, Q_{\lambda} = \sum_{\Box \in \text{Rem}(\lambda)}  (\gamma_{\Box})^k \, (\delta_{i_{\Box}, j_{\Box}} - 2)  \, Q_{\lambda - \Box} \,.
	\end{equation}
	Operators $\hat{\Psi}_a$ are defined by commutators
	\begin{equation}
		\hat{\Psi}_{a+b} = \Big[ \hat{E}_a, \hat{F}_b \Big]
	\end{equation}
	and form a  commutative family
	\begin{equation}
		\Big[ \hat{\Psi}_a, \hat{\Psi}_b \Big] = 0.
	\end{equation}
	The initial operator $\hat{U}$ is included in the  commutative family
	\begin{equation}
		72 \hat{U} = 1 - 2 \hat{\Psi}_1 + \hat{\Psi}_2,
	\end{equation}
	therefore Q-Schur polynomials are common eigenfunctions of $\hat{\Psi}_a$. We observe the following relations on differential operators $\hat{E}_k, \hat{F}_k, \hat{\Psi}_k$
	\begin{align}
		\begin{aligned}
			\Big[ \hat{E}_{n+3}, \hat{E}_{m} \Big] - 3 \Big[ \hat{E}_{n+2}, \hat{E}_{m+1} \Big] + 3 \Big[ \hat{E}_{n+1}, \hat{E}_{m+2} \Big] - \Big[ \hat{E}_{n}, \hat{E}_{m+3} \Big] = \\
			=+ 6 \Big[ \hat{E}_{n+2}, \hat{E}_{m} \Big] - 6 \Big[ \hat{E}_{n}, \hat{E}_{m+2} \Big] - 12 \Big[ \hat{E}_{n+1}, \hat{E}_{m} \Big]+ 12 \Big[ \hat{E}_{n}, \hat{E}_{m+1} \Big],
		\end{aligned}
	\end{align}
	\begin{align}
		\begin{aligned}
			\Big[ \hat{F}_{n+3}, \hat{F}_{m} \Big] - 3 \Big[ \hat{F}_{n+2}, \hat{F}_{m+1} \Big] + 3 \Big[ \hat{F}_{n+1}, \hat{F}_{m+2} \Big] - \Big[ \hat{F}_{n}, \hat{F}_{m+3} \Big] = \\
			=+6 \Big[ \hat{F}_{n+2}, \hat{F}_{m} \Big] + 6 \Big[ \hat{F}_{n}, \hat{F}_{m+2} \Big] - 12 \Big[ \hat{F}_{n+1}, \hat{F}_{m} \Big]+ 12 \Big[ \hat{F}_{n}, \hat{F}_{m+1} \Big],
		\end{aligned}
	\end{align}
	\begin{align}
		\begin{aligned}
			\Big[ \hat{\Psi}_{n+3}, \hat{E}_{m} \Big] - 3 \Big[ \hat{\Psi}_{n+2}, \hat{E}_{m+1} \Big] + 3 \Big[ \hat{\Psi}_{n+1}, \hat{E}_{m+2} \Big] - \Big[ \hat{\Psi}_{n}, \hat{E}_{m+3} \Big] = \\
			=+ 6 \Big[ \hat{\Psi}_{n+2}, \hat{E}_{m} \Big] - 6 \Big[ \hat{\Psi}_{n}, \hat{E}_{m+2} \Big] - 12 \Big[ \hat{\Psi}_{n+1}, \hat{E}_{m} \Big]+ 12 \Big[ \hat{\Psi}_{n}, \hat{E}_{m+1} \Big],
		\end{aligned}
	\end{align}
	\begin{align}
		\begin{aligned}
			\label{psi f Q rel}
			\Big[ \hat{\Psi}_{n+3}, \hat{F}_{m} \Big] - 3 \Big[ \hat{\Psi}_{n+2}, \hat{F}_{m+1} \Big] + 3 \Big[ \hat{\Psi}_{n+1}, \hat{F}_{m+2} \Big] - \Big[ \hat{\Psi}_{n}, \hat{F}_{m+3} \Big] = \\
			=+ 6 \Big[ \hat{\Psi}_{n+2}, \hat{F}_{m} \Big] - 6 \Big[ \hat{\Psi}_{n}, \hat{F}_{m+2} \Big] - 12 \Big[ \hat{\Psi}_{n+1}, \hat{F}_{m} \Big] + 12 \Big[ \hat{\Psi}_{n}, \hat{F}_{m+1} \Big].
		\end{aligned}
	\end{align}
	Note the similarity with the Schur case -- the coefficients in front of the commutators are always the same for these four relations. Of course, we verify these relations in specific representation of the algebra and the true relations may differ or there may by more relations. In addition to quadratic relations presented above, we observe quartic relations of the following form
	\begin{equation}
		\text{Sym}_{i,j,k,l} \Big[ \hat{E}_i, \Big[ \hat{E}_{j}, \Big[ \hat{E}_k, \hat{E}_{l+1} \Big] \Big] \Big] = 0,
	\end{equation}
	\begin{equation}
		\text{Sym}_{i,j,k,l} \Big[ \hat{F}_i, \Big[ \hat{F}_{j}, \Big[ \hat{F}_k, \hat{F}_{l+1} \Big] \Big] \Big] = 0,
	\end{equation}
	that are similar to Serre relations \eqref{serre rel}.
	\subsection{Commutative integer rays of operators}
	\label{sec:: Q Schur comm rays}
	In Q-Schur case we observe commutative subalgebras $\hat{\mathcal{H}}_{n}^{(M)}$, enumerated by integer numbers $M = 0,1,2,\ldots$
	\begin{equation}
		\Big[ \hat{\mathcal{H}}_{n}^{(M)}, \hat{\mathcal{H}}_{m}^{(M)} \Big] = 0,
	\end{equation}
	that are generalization of those considered in Sec.\ref{sec:: Schur comm integer rays}. The first commutative family $M=0$ correspond to variables $\hat{\mathcal{H}}_{n}^{(0)} = (-1)^n 6^n (2n+1)!! \cdot \tau_n$. Explicit examples from small level read
	\begin{align}
		\begin{aligned}
			\hat{\mathcal{H}}_{0}^{(0)} &= \hat{E}_0 = \tau_0, \\
			\hat{\mathcal{H}}_{1}^{(0)} &= \Big[ \hat{E}_0, \Big[ \hat{E}_1, \hat{E}_0 \Big] \Big] =  - 6 \cdot 3 \cdot \tau_1, \\
			\hat{\mathcal{H}}_{2}^{(0)} &= \Big[ \hat{E}_0, \Big[ \hat{E}_1, \Big[ \hat{E}_0, \Big[ \hat{E}_1, \hat{E}_0 \Big] \Big] \Big] \Big] = 6^2 \cdot 3 \cdot 5 \cdot \tau_2, \\
			\hat{\mathcal{H}}_{3}^{(0)} &= \Big[ \hat{E}_0, \Big[ \hat{E}_1, \Big[ \hat{E}_0, \Big[ \hat{E}_1, \Big[ \hat{E}_0, \Big[ \hat{E}_1, \hat{E}_0 \Big] \Big] \Big] \Big] \Big] \Big] = -6^3 \cdot 3 \cdot 5 \cdot 7 \cdot \tau_3, \\
			\ldots
		\end{aligned}
	\end{align}
	The recursive relation takes the following form
	\begin{equation}
		\hat{\mathcal{H}}_{n}^{(0)} = \Big[ \hat{E}_0, \Big[ \hat{E}_1, \hat{\mathcal{H}}_{n-1}^{(0)} \Big] \Big].
	\end{equation}
	Dual commutative families $\hat{\mathcal{H}}_{m}^{*(M)}$ 
	\begin{equation}
		\Big[ \hat{\mathcal{H}}_{n}^{*(M)}, \hat{\mathcal{H}}_{m}^{*(M)} \Big] = 0,
	\end{equation}
	are constructed with the help of operators $\hat{F}_k$. $M=0$ case corresponds to derivatives $\hat{\mathcal{H}}_{n}^{*(0)} = (-1)^{n+1} (24)^n (2n-1)!! \cdot \frac{\partial}{\partial \tau_n}$, and this fact can be seen from small levels
	\begin{align}
		\begin{aligned}
			\hat{\mathcal{H}}_{0}^{*(0)} &= \hat{F}_0 = -\frac{\partial}{\partial \tau_0}, \\
			\hat{\mathcal{H}}_{1}^{*(0)} &= \Big[ \hat{F}_0, \Big[ \hat{F}_1, \hat{F}_0 \Big] \Big] =  24 \cdot \frac{\partial}{\partial \tau_1}, \\
			\hat{\mathcal{H}}_{2}^{*(0)} &= \Big[ \hat{F}_0, \Big[ \hat{F}_1, \Big[ \hat{F}_0, \Big[ \hat{F}_1, \hat{F}_0 \Big] \Big] \Big] \Big] = -(24)^2 \cdot 3 \cdot \frac{\partial}{\partial \tau_2}, \\
			\hat{\mathcal{H}}_{3}^{*(0)} &= \Big[ \hat{F}_0, \Big[ \hat{F}_1, \Big[ \hat{F}_0, \Big[ \hat{F}_1, \Big[ \hat{F}_0, \Big[ \hat{F}_1, \hat{F}_0 \Big] \Big] \Big] \Big] \Big] \Big] = (24)^3 \cdot 3 \cdot 5 \cdot \frac{\partial}{\partial \tau_3}, \\
			\ldots
		\end{aligned}
	\end{align}
	The recursive relation follows
	\begin{equation}
		\hat{\mathcal{H}}_{n}^{*(0)} = \Big[ \hat{F}_0, \Big[ \hat{F}_1, \hat{\mathcal{H}}_{n-1}^{*(0)} \Big] \Big].
	\end{equation}
	The higher commutative families are constructed via the following recursive relations
	\begin{tcolorbox}
		\begin{align}
			\begin{aligned}
				\hat{\mathcal{H}}_{n}^{(M)} &= \Big[ \hat{E}_{M}, \Big[ \hat{E}_{M+1}, \hat{\mathcal{H}}_{n-1}^{(M)} \Big] \Big], &\hspace{20mm} \hat{\mathcal{H}}_{0}^{(M)} &= \hat{E}_{M}, \\
				\hat{\mathcal{H}}_{n}^{*(M)} &= \Big[ \hat{F}_{M}, \Big[ \hat{F}_{M+1}, \hat{\mathcal{H}}_{n-1}^{*(M)} \Big] \Big], &\hspace{20mm} \hat{\mathcal{H}}_{0}^{*(M)} &= \hat{F}_{M}. 
			\end{aligned}
		\end{align}
	\end{tcolorbox}
	\subsection{Q-Schur polynomials from the first commutative family}
	\label{sec:: Q Schurs from the first family}
	Similarly to Sec.\ref{sec::Schur from first family} one can compute Q-Schur polynomials from the following relations
	\begin{equation}
		\hat{E}_0 \, Q_{\lambda} = \sum_{\Box \in \text{Add}(\lambda)} Q_{\lambda + \Box} \,, 
	\end{equation}
	\begin{equation}
		\hat{E}_1 \, Q_{\lambda} = \sum_{\Box \in \text{Add}(\lambda)} \gamma_{\Box} \, Q_{\lambda + \Box} \,, 
	\end{equation}
	where $\gamma_{\Box} = (j_{\Box} - i_{\Box}+1)^3 - (j_{\Box} - i_{\Box})^3$. It is not required to know explicit form of operator $\hat{E}_1$ only its action on Q-Schur polynomials. Then using relation
	\begin{equation}
		\hat{\mathcal{H}}_{n}^{(0)} = (-1)^n 6^n (2n+1)!! \cdot \tau_n
	\end{equation}
	and starting polynomial $Q_{\varnothing} = 1$, one can compute Q-Schur polynomials. We provide examples from small levels. 
	
	1 level:
	\begin{align}
		\begin{aligned}
			\tau_0 \cdot 1 &= \hat{E}_0 \, Q_{\varnothing} = Q_{[1]}  \\
		\end{aligned}
	\end{align}
	
	2 level:
	\begin{align}
		\begin{aligned}
			\tau_0^2 \cdot 1 &= \hat{E}_0 \hat{E}_0 \, Q_{\varnothing} = Q_{[2]}  \\
		\end{aligned}
	\end{align}

	3 level:
	\begin{align}
		\begin{aligned}
			\tau_0^3 \cdot 1 &= \hat{E}_0 \hat{E}_0 \hat{E}_0 \, Q_{\varnothing} = Q_{[3]} + Q_{[2,1]},  \\
			\tau_1 \cdot 1 &=  -\frac{1}{18} \Big[ \hat{E}_0, \Big[ \hat{E}_1, \hat{E}_0 \Big] \Big] \, Q_{\varnothing} = \frac{1}{3} \left( Q_{[3]} - 2 Q_{[2,1]} \right),
		\end{aligned}
	\end{align}
	\begin{equation}
		Q_{[3]} = \frac{2 \tau _0^3}{3}+\tau _1, \hspace{20mm} Q_{[2,1]} = \frac{\tau _0^3}{3}-\tau _1.
	\end{equation}

	4 level:
	\begin{align}
		\begin{aligned}
			\tau_0^4 \cdot 1 &= \hat{E}_0 \hat{E}_0 \hat{E}_0 \hat{E}_0 \, Q_{\varnothing} = Q_{[4]} + 2Q_{[3,1]},  \\
			\tau_0 \tau_1 \cdot 1 &=  -\frac{1}{18} \hat{E}_0 \Big[ \hat{E}_0, \Big[ \hat{E}_1, \hat{E}_0 \Big] \Big] \, Q_{\varnothing} = \frac{1}{3} \left( Q_{[4]} - Q_{[3,1]} \right),
		\end{aligned}
	\end{align}
	\begin{equation}
		Q_{[4]} = \frac{\tau _0^4}{3}+2 \tau _0 \tau _1, \hspace{20mm} Q_{[3,1]} = \frac{\tau _0^4}{3}-\tau _0 \tau _1.
	\end{equation}

	5 level:
	\begin{align}
		\begin{aligned}
			\tau_0^5 \cdot 1 &= \hat{E}_0 \hat{E}_0 \hat{E}_0 \hat{E}_0 \hat{E}_0 \, Q_{\varnothing} = Q_{[5]} + 3Q_{[4,1]} + 2 Q_{[2,2]},  \\
			\tau_0^2 \tau_1 \cdot 1 &=  -\frac{1}{18} \hat{E}_0 \hat{E}_0 \Big[ \hat{E}_0, \Big[ \hat{E}_1, \hat{E}_0 \Big] \Big] \, Q_{\varnothing} = \frac{1}{3} \left( Q_{[5]} - Q_{[3,2]} \right), \\
			\tau_2 \cdot 1 &=  \frac{1}{540} \Big[ \hat{E}_0, \Big[ \hat{E}_1,\Big[ \hat{E}_0, \Big[ \hat{E}_1, \hat{E}_0 \Big] \Big] \Big] \Big] \, Q_{\varnothing} = \frac{1}{5} \left( Q_{[5]} - 2Q_{[4,1]} + 2 Q_{[3,2]} \right),
		\end{aligned}
	\end{align}
	\begin{equation}
		Q_{[5]} = \frac{2 \tau _0^5}{15}+2 \tau _0^2 \tau _1+\tau _2, \hspace{15mm} Q_{[4,1]} = \frac{\tau _0^5}{5}-\tau _2, \hspace{15mm} Q_{[3,2]} = \frac{2 \tau _0^5}{15}-\tau _0^2 \tau _1+\tau _2.
	\end{equation}

	\subsection{Explicit form of operators $\hat{\Psi}_a$}
	\label{sec:: Q Schur Casimirs}
	All operators $\hat{\Psi}_a$ in the Q-Schur case have \textit{box additivity} property just like operators in the Schur case. In other words, all operators $\hat{\Psi}_a$ obey relation of the following from
	\begin{equation}
		\hat{\Psi}_{\Omega} \, Q_{\lambda} = \left( \sum_{\Box \in \lambda} \Omega( j_{\Box} - i_{\Box} )\right) Q_{\lambda},
	\end{equation}
	with the proper choice of the function $\Omega(x)$. We expand this operator in terms of Q-Schur polynomials and corresponding dual operators
	\begin{equation}
		\label{Q Schur expansion}
		\hat{\Psi}_{\Omega} = \sum_{\lambda, \mu} A_{\mu, \nu} \, Q_{\lambda} \hat{Q}_{\mu},
	\end{equation}
	where $A_{\mu, \nu}$ are constants and the sum runs over strict partitions $\mu$, $\nu$ such that $|\mu| = |\nu|$. Dual operators are defined by the following rule
	\begin{equation}
		\hat{Q}_{\lambda} := Q_{\lambda} \left( \tau_k \to \frac{4^k}{2k+1} \frac{\partial}{\partial \tau_k}\right).
	\end{equation}
	Expansion of operators \eqref{Q Schur expansion} with box additivity property involve only \textit{two row} diagrams -- similarly to Schur case and single hook diagrams. Note that single row diagram $[n] = [n,0]$ also can be represented as two row diagram. The resulting formula has the following form
	\begin{tcolorbox}
		\begin{equation}
			\hat{\Psi}_{\Omega} = \sum_{n=1}^{\infty} \sum_{i,j = 0}^{\text{IntPart}(\frac{n-1}{2})} (-2)^{1-n}(2 - \delta_{i,0}) (2 - \delta_{j,0}) \left[ \sum_{k = |j-i|}^{n - 1 - i - j} (-1)^k \,  \Omega(k) \right] Q_{[n-i,i]} \hat{Q}_{[n-j,j]}
		\end{equation}
	\end{tcolorbox}
	Therefore the form of $\hat{\Psi}_a$ operators is fully fixed by functions $\Omega_a(x)$. All operators have constant term $1$, but it does not affect the expansion and we omit it. We provide several explicit examples
	\begin{align}
		\begin{aligned}
			\Omega_0(x) &= 0, \\
			\Omega_1(x) &= 12, \\
			\Omega_2(x) &= 96+216 x+216 x^2, \\
			\Omega_3(x) &= 684+2592 x+4212 x^2+3240 x^3+1620 x^4, \\
			\Omega_4(x) &= 4800+24624 x+55728 x^2+71280 x^3+58320 x^4+27216 x^5+9072 x^6, \\
			\ldots
		\end{aligned}
	\end{align}
	The recursive relation on functions $\Omega_a(x)$ follows from the relations (see Sec.\ref{sec::towards the algebra}) 
	\begin{align}
		\begin{aligned}
			\Omega_{a+3}(x)-3 \left(\gamma + 2\right) \cdot \Omega_{a+2}(x) +3(\gamma^2+4) \Omega_{a+1}(x) - \gamma(\gamma^2 +12) \cdot \Omega_{a}(x) = 0,
		\end{aligned}
	\end{align}
	where $\gamma = (x+1)^3-x^3$. The characteristic polynomial has three roots
	\begin{align}
		\begin{aligned}
			t^{a+3}-3 \left(\gamma + 2\right) \cdot t^{a+2} +3(\gamma^2+4) t^{a+1} - \gamma(\gamma^2 +12) \cdot t^{a} = \\
			= - t^a \left(1-3 x+3 x^2-t \right) \left(1+3 x+3 x^2-t \right) \left(7+9 x+3 x^2-t \right).
		\end{aligned}
	\end{align}
	Therefore the final answer takes the following form
	\begin{equation}
		\boxed{\boxed{
		\Omega_a(x) = 2\left(1-3 x+3 x^2 \right)^a +  2 \left(7+9 x+3 x^2\right)^a -4 \left(1+3 x+3 x^2 \right)^a.
	}}
	\end{equation}
	
	\section*{Acknowledgements}
	We are grateful for inspiring discussions to L.Bishler, Ya.Drachov,  A.D.Mironov, A.Oreshina and A.Popolitov.

	We gratefully acknowledge support from the Ministry of Science and Higher Education of the Russian Federation (agreement no. 075-03-2025-662). This work is partly supported by the grants of the Foundation for the Advancement of Theoretical Physics and Mathematics “BASIS”.

	\bibliographystyle{utphys}
	\bibliography{main}

\providecommand{\href}[2]{#2}\begingroup\raggedright\begin{thebibliography}{10}

\bibitem{Drinfeld:1985rx}
V.~G. Drinfeld, ``{Hopf algebras and the quantum Yang-Baxter equation},'' {\em
  Sov. Math. Dokl.} {\bfseries 32} (1985) 254--258.

\bibitem{Faddeev:1996iy}
L.~D. Faddeev, ``{How algebraic Bethe ansatz works for integrable model},'' in
  {\em {Les Houches School of Physics: Astrophysical Sources of Gravitational
  Radiation}}, pp.~pp. 149--219.
\newblock 5, 1996.
\newblock \href{http://arxiv.org/abs/hep-th/9605187}{{\ttfamily
  arXiv:hep-th/9605187}}.

\bibitem{Dolan:2004ps}
L.~Dolan, C.~R. Nappi, and E.~Witten,
  \href{http://dx.doi.org/10.1142/9789812702340_0036}{``{Yangian symmetry in D
  = 4 superconformal Yang-Mills theory},''} in {\em {3rd International
  Symposium on Quantum Theory and Symmetries}}, pp.~300--315.
\newblock 1, 2004.
\newblock \href{http://arxiv.org/abs/hep-th/0401243}{{\ttfamily
  arXiv:hep-th/0401243}}.

\bibitem{Tsymbaliuk:2014fvq}
A.~Tsymbaliuk, ``{The affine Yangian of $\mathfrak{gl}_1$ revisited},''
  \href{http://dx.doi.org/10.1016/j.aim.2016.08.041}{{\em Adv. Math.}
  {\bfseries 304} (2017) 583--645},
  \href{http://arxiv.org/abs/1404.5240}{{\ttfamily arXiv:1404.5240 [math.RT]}}.

\bibitem{Okounkov:2015spn}
A.~Okounkov, ``{Lectures on K-theoretic computations in enumerative
  geometry},'' \href{http://arxiv.org/abs/1512.07363}{{\ttfamily
  arXiv:1512.07363 [math.AG]}}.

\bibitem{Prochazka:2015deb}
T.~Proch\'azka, ``{$ \mathcal{W} $ -symmetry, topological vertex and affine
  Yangian},'' \href{http://dx.doi.org/10.1007/JHEP10(2016)077}{{\em JHEP}
  {\bfseries 10} (2016) 077}, \href{http://arxiv.org/abs/1512.07178}{{\ttfamily
  arXiv:1512.07178 [hep-th]}}.

\bibitem{Ding:1996mq}
J.~Ding, J.-t. Ding, and K.~Iohara, ``{Generalization and deformation of
  Drinfeld quantum affine algebras},''
  \href{http://dx.doi.org/10.1023/A:1007341410987}{{\em Lett. Math. Phys.}
  {\bfseries 41} (1997) 181--193},
  \href{http://arxiv.org/abs/q-alg/9608002}{{\ttfamily arXiv:q-alg/9608002}}.

\bibitem{Miki:2007mer}
K.~Miki, ``{A (q,\ensuremath{\gamma}) analog of the W1+\ensuremath{\infty}
  algebra},'' \href{http://dx.doi.org/10.1063/1.2823979}{{\em J. Math. Phys.}
  {\bfseries 48} no.~12, (2007) 123520}.

\bibitem{Mironov:2016yue}
A.~Mironov, A.~Morozov, and Y.~Zenkevich, ``{Ding-Iohara-Miki symmetry of
  network matrix models},'' \href{http://arxiv.org/abs/1603.05467}{{\ttfamily
  arXiv:1603.05467 [hep-th]}}.

\bibitem{Awata:2016riz}
H.~Awata, H.~Kanno, T.~Matsumoto, A.~Mironov, A.~Morozov, A.~Morozov,
  Y.~Ohkubo, and Y.~Zenkevich, ``{Explicit examples of DIM constraints for
  network matrix models},''
  \href{http://dx.doi.org/10.1007/JHEP07(2016)103}{{\em JHEP} {\bfseries 07}
  (2016) 103}, \href{http://arxiv.org/abs/1604.08366}{{\ttfamily
  arXiv:1604.08366 [hep-th]}}.

\bibitem{Feigin2}
B.~Feigin, M.~Jimbo, T.~Miwa, and E.~Mukhin, ``Quantum toroidal gl1-algebra:
  Plane partitions,'' \href{http://dx.doi.org/10.1215/21562261-1625217}{{\em
  Kyoto Journal of Mathematics} {\bfseries 52} no.~3, (2012) 621 -- 659}.

\bibitem{Galakhov:2020vyb}
D.~Galakhov and M.~Yamazaki, ``{Quiver Yangian and Supersymmetric Quantum
  Mechanics},'' \href{http://dx.doi.org/10.1007/s00220-022-04490-y}{{\em
  Commun. Math. Phys.} {\bfseries 396} no.~2, (2022) 713--785},
  \href{http://arxiv.org/abs/2008.07006}{{\ttfamily arXiv:2008.07006
  [hep-th]}}.

\bibitem{Mironov:2019uoy}
A.~Mironov and A.~Morozov, ``{On generalized Macdonald polynomials},''
  \href{http://dx.doi.org/10.1007/JHEP01(2020)110}{{\em JHEP} {\bfseries 01}
  (2020) 110}, \href{http://arxiv.org/abs/1907.05410}{{\ttfamily
  arXiv:1907.05410 [hep-th]}}.

\bibitem{Mironov:2020pcd}
A.~Mironov, V.~Mishnyakov, A.~Morozov, and A.~Popolitov, ``{Commutative
  families in $W_{\infty}$, integrable many-body systems and hypergeometric
  \ensuremath{\tau}-functions},''
  \href{http://dx.doi.org/10.1007/JHEP09(2023)065}{{\em JHEP} {\bfseries 23}
  (2020) 065}, \href{http://arxiv.org/abs/2306.06623}{{\ttfamily
  arXiv:2306.06623 [hep-th]}}.

\bibitem{Galakhov:2023mak}
D.~Galakhov, A.~Morozov, and N.~Tselousov, ``{Super-Schur polynomials for
  Affine Super Yangian Y($ \hat{\mathfrak{gl}} _{1|1}$)},''
  \href{http://dx.doi.org/10.1007/JHEP08(2023)049}{{\em JHEP} {\bfseries 08}
  (2023) 049}, \href{http://arxiv.org/abs/2307.03150}{{\ttfamily
  arXiv:2307.03150 [hep-th]}}.

\bibitem{Galakhov:2024mbz}
D.~Galakhov, A.~Morozov, and N.~Tselousov, ``{Simple representations of BPS
  algebras: the case of $Y(\widehat{\mathfrak {gl}}_2)$},''
  \href{http://dx.doi.org/10.1140/epjc/s10052-024-12952-x}{{\em Eur. Phys. J.
  C} {\bfseries 84} no.~6, (2024) 604},
  \href{http://arxiv.org/abs/2402.05920}{{\ttfamily arXiv:2402.05920
  [hep-th]}}.

\bibitem{Mironov:2024sbc}
A.~Mironov, A.~Morozov, and A.~Popolitov, ``{Commutative families in DIM
  algebra, integrable many-body systems and q, t matrix models},''
  \href{http://dx.doi.org/10.1007/JHEP09(2024)200}{{\em JHEP} {\bfseries 09}
  (2024) 200}, \href{http://arxiv.org/abs/2406.16688}{{\ttfamily
  arXiv:2406.16688 [hep-th]}}.

\bibitem{Macdonald}
I.~G. Macdonald, {\em Symmetric Functions and Hall Polynomials}.
\newblock Oxford Mathematical Monographs, 1998.

\bibitem{Yamazaki:2010fz}
M.~Yamazaki, ``{Crystal Melting and Wall Crossing Phenomena},''
  \href{http://dx.doi.org/10.1142/S0217751X11051482}{{\em Int. J. Mod. Phys. A}
  {\bfseries 26} (2011) 1097--1228},
  \href{http://arxiv.org/abs/1002.1709}{{\ttfamily arXiv:1002.1709 [hep-th]}}.

\bibitem{Galakhov:2021xum}
D.~Galakhov, W.~Li, and M.~Yamazaki, ``{Shifted quiver Yangians and
  representations from BPS crystals},''
  \href{http://dx.doi.org/10.1007/JHEP08(2021)146}{{\em JHEP} {\bfseries 08}
  (2021) 146}, \href{http://arxiv.org/abs/2106.01230}{{\ttfamily
  arXiv:2106.01230 [hep-th]}}.

\bibitem{NW}
G.~Noshita and A.~Watanabe, ``{Shifted quiver quantum toroidal algebra and
  subcrystal representations},''
  \href{http://dx.doi.org/10.1007/JHEP05(2022)122}{{\em JHEP} {\bfseries 05}
  (2022) 122}, \href{http://arxiv.org/abs/2109.02045}{{\ttfamily
  arXiv:2109.02045 [hep-th]}}.

\bibitem{Galakhov:2025phf}
D.~Galakhov, A.~Morozov, and N.~Tselousov, ``{Super-Hamiltonians for
  super-Macdonald polynomials},''
  \href{http://arxiv.org/abs/2501.14714}{{\ttfamily arXiv:2501.14714
  [hep-th]}}.

\bibitem{PSR}
C.~Pope, X.~Shen, and L.~Romans, ``$W_{\infty}$ and the Racah-Wigner algebra,''
  {\em Nuclear Physics B} {\bfseries 339} no.~1, (1990) 191--221.

\bibitem{Awata:1994tf}
H.~Awata, M.~Fukuma, Y.~Matsuo, and S.~Odake, ``{Representation theory of the
  $W_{1+\infty}$ algebra},'' \href{http://dx.doi.org/10.1143/PTPS.118.343}{{\em
  Prog. Theor. Phys. Suppl.} {\bfseries 118} (1995) 343--374},
  \href{http://arxiv.org/abs/hep-th/9408158}{{\ttfamily arXiv:hep-th/9408158}}.

\bibitem{Jimbo:1983if}
M.~Jimbo and T.~Miwa, ``{Solitons and Infinite Dimensional Lie Algebras},''
  \href{http://dx.doi.org/10.2977/prims/1195182017}{{\em Publ. Res. Inst. Math.
  Sci. Kyoto} {\bfseries 19} (1983) 943}.

\bibitem{Drinfeld:1987sy}
V.~G. Drinfeld, ``{A New realization of Yangians and quantized affine
  algebras},'' {\em Sov. Math. Dokl.} {\bfseries 36} (1988) 212--216.

\bibitem{Mironov:2023zwi}
A.~Mironov and A.~Morozov, ``{Many-body integrable systems implied by WLZZ
  models},'' \href{http://dx.doi.org/10.1016/j.physletb.2023.137964}{{\em Phys.
  Lett. B} {\bfseries 842} (2023) 137964},
  \href{http://arxiv.org/abs/2303.05273}{{\ttfamily arXiv:2303.05273
  [hep-th]}}.

\bibitem{Mironov:2019mah}
A.~Mironov and A.~Morozov, ``{Hook variables: Cut-and-join operators and $\tau$
  -functions},'' \href{http://dx.doi.org/10.1016/j.physletb.2020.135362}{{\em
  Phys. Lett. B} {\bfseries 804} (2020) 135362},
  \href{http://arxiv.org/abs/1912.00635}{{\ttfamily arXiv:1912.00635
  [hep-th]}}.

\bibitem{Mironov:2021taq}
A.~Mironov, A.~Morozov, and A.~Zhabin, ``{Connection between cut-and-join and
  Casimir operators},''
  \href{http://dx.doi.org/10.1016/j.physletb.2021.136668}{{\em Phys. Lett. B}
  {\bfseries 822} (2021) 136668},
  \href{http://arxiv.org/abs/2105.10978}{{\ttfamily arXiv:2105.10978
  [hep-th]}}.

\bibitem{Mironov:2021wfh}
A.~Mironov, A.~Morozov, and A.~Zhabin, ``{Spin Hurwitz theory and Miwa
  transform for the Schur Q-functions},''
  \href{http://dx.doi.org/10.1016/j.physletb.2022.137131}{{\em Phys. Lett. B}
  {\bfseries 829} (2022) 137131},
  \href{http://arxiv.org/abs/2111.05776}{{\ttfamily arXiv:2111.05776
  [hep-th]}}.

\bibitem{Ker1Mironov:2019}
A.~Mironov and A.~Morozov, ``{On Hamiltonians for Kerov functions},''
  \href{http://dx.doi.org/10.1140/epjc/s10052-020-7811-3}{{\em Eur. Phys. J. C}
  {\bfseries 80} no.~3, (2020) 277},
  \href{http://arxiv.org/abs/1908.05176}{{\ttfamily arXiv:1908.05176
  [hep-th]}}.

\bibitem{Mironov:2023wga}
A.~Mironov, V.~Mishnyakov, A.~Morozov, and A.~Popolitov, ``{Commutative
  subalgebras from Serre relations},''
  \href{http://dx.doi.org/10.1016/j.physletb.2023.138122}{{\em Phys. Lett. B}
  {\bfseries 845} (2023) 138122},
  \href{http://arxiv.org/abs/2307.01048}{{\ttfamily arXiv:2307.01048
  [hep-th]}}.

\bibitem{MMintsystWLZZ}
A.~Mironov and A.~Morozov, ``{Many-body integrable systems implied by WLZZ
  models},'' \href{http://arxiv.org/abs/2303.05273}{{\ttfamily arXiv:2303.05273
  [hep-th]}}.

\bibitem{DATE1982343}
E.~Date, M.~Jimbo, M.~Kashiwara, and T.~Miwa, ``Transformation groups for
  soliton equations: IV. A new hierarchy of soliton equations of KP-type,''
  \href{http://dx.doi.org/https://doi.org/10.1016/0167-2789(82)90041-0}{{\em
  Physica D: Nonlinear Phenomena} {\bfseries 4} no.~3, (1982) 343--365}.
  \url{https://www.sciencedirect.com/science/article/pii/0167278982900410}.

\bibitem{Alexandrov:2020yxf}
A.~Alexandrov, ``{Intersection numbers on $ {\overline{M}}_{g,n} $ and BKP
  hierarchy},'' \href{http://dx.doi.org/10.1007/JHEP09(2021)013}{{\em JHEP}
  {\bfseries 09} (2021) 013}, \href{http://arxiv.org/abs/2012.07573}{{\ttfamily
  arXiv:2012.07573 [math-ph]}}.

\bibitem{Alexandrov:2020nzt}
A.~Alexandrov, ``{KdV solves BKP},''
  \href{http://dx.doi.org/10.1073/pnas.2101917118}{{\em Proc. Nat. Acad. Sci.}
  {\bfseries 118} (2021) e2101917118},
  \href{http://arxiv.org/abs/2012.10448}{{\ttfamily arXiv:2012.10448
  [nlin.SI]}}.

\bibitem{Drachov:2023xyz}
Y.~Drachov and A.~Zhabin, ``{Genus expansion of matrix models and $\hbar $
  expansion of BKP hierarchy},''
  \href{http://dx.doi.org/10.1140/epjc/s10052-023-11617-5}{{\em Eur. Phys. J.
  C} {\bfseries 83} no.~5, (2023) 437},
  \href{http://arxiv.org/abs/2302.03949}{{\ttfamily arXiv:2302.03949
  [hep-th]}}.

\bibitem{MThunt}
A.~Morozov and N.~Tselousov, ``{Hunt for 3-Schur polynomials},''
  \href{http://dx.doi.org/10.1016/j.physletb.2023.137887}{{\em Phys. Lett. B}
  {\bfseries 840} (2023) 137887},
  \href{http://arxiv.org/abs/2211.14956}{{\ttfamily arXiv:2211.14956
  [hep-th]}}.

\bibitem{Morozov:2023vra}
A.~Morozov and N.~Tselousov, ``{3-Schurs from explicit representation of
  Yangian $ \textrm{Y}\left({\hat{\mathfrak{gl}}}_1\right) $. Levels
  1\textendash{}5},'' \href{http://dx.doi.org/10.1007/JHEP11(2023)165}{{\em
  JHEP} {\bfseries 11} (2023) 165},
  \href{http://arxiv.org/abs/2305.12282}{{\ttfamily arXiv:2305.12282
  [hep-th]}}.

\bibitem{Wang1}
N.~Wang, ``{Affine Yangian and 3-Schur functions},''
  \href{http://dx.doi.org/10.1016/j.nuclphysb.2020.115173}{{\em Nucl. Phys. B}
  {\bfseries 960} (2020) 115173}.

\bibitem{Morozov:2018fga}
A.~Y. Morozov, ``{Cut-and-join operators and Macdonald polynomials from the
  3-Schur functions},'' \href{http://arxiv.org/abs/1810.00395}{{\ttfamily
  arXiv:1810.00395 [hep-th]}}.

\bibitem{Qschurs1}
A.~D. Mironov and A.~Morozov, ``{Generalized Q-functions for GKM},''
  \href{http://dx.doi.org/10.1016/j.physletb.2021.136474}{{\em Phys. Lett. B}
  {\bfseries 819} (2021) 136474},
  \href{http://arxiv.org/abs/2101.08759}{{\ttfamily arXiv:2101.08759
  [hep-th]}}.

\bibitem{Morozov:2020ccp}
A.~Morozov, ``{A new kind of anomaly: on W-constraints for GKM},''
  \href{http://dx.doi.org/10.1007/JHEP10(2021)213}{{\em JHEP} {\bfseries 21}
  (2020) 213}, \href{http://arxiv.org/abs/2108.07198}{{\ttfamily
  arXiv:2108.07198 [hep-th]}}.

\bibitem{MMN}
A.~Mironov, A.~Morozov, and S.~Natanzon, ``{Complete Set of Cut-and-Join
  Operators in Hurwitz-Kontsevich Theory},''
  \href{http://arxiv.org/abs/0904.4227}{{\ttfamily arXiv:0904.4227 [hep-th]}}.

\bibitem{Wang:2022fxr}
R.~Wang, F.~Liu, C.-H. Zhang, and W.-Z. Zhao, ``{Superintegrability for ($\beta
  $-deformed) partition function hierarchies with W-representations},''
  \href{http://dx.doi.org/10.1140/epjc/s10052-022-10875-z}{{\em Eur. Phys. J.
  C} {\bfseries 82} no.~10, (2022) 902},
  \href{http://arxiv.org/abs/2206.13038}{{\ttfamily arXiv:2206.13038
  [hep-th]}}.

\bibitem{MMMPWZ}
A.~Mironov, V.~Mishnyakov, A.~Morozov, A.~Popolitov, R.~Wang, and W.-Z. Zhao,
  ``{Interpolating matrix models for WLZZ series},''
  \href{http://arxiv.org/abs/2301.04107}{{\ttfamily arXiv:2301.04107
  [hep-th]}}.

\bibitem{MMMPZ}
A.~Mironov, V.~Mishnyakov, A.~Morozov, A.~Popolitov, and W.-Z. Zhao, ``{On
  KP-integrable skew Hurwitz \ensuremath{\tau}-functions and their
  \ensuremath{\beta}-deformations},''
  \href{http://arxiv.org/abs/2301.11877}{{\ttfamily arXiv:2301.11877}}.

\bibitem{Azheev:2025wti}
B.~Azheev and N.~Tselousov, ``{Towards construction of superintegrable basis in
  matrix models},''
  \href{http://dx.doi.org/10.1016/j.nuclphysb.2025.116975}{{\em Nucl. Phys. B}
  {\bfseries 1018} (2025) 116975},
  \href{http://arxiv.org/abs/2503.07583}{{\ttfamily arXiv:2503.07583
  [hep-th]}}.

\end{thebibliography}\endgroup

\end{document}